\documentclass[journal]{IEEEtran}
\usepackage{amsmath,amsfonts}
\usepackage{algorithm}
\usepackage[noend]{algorithmic}
\usepackage{array}
\usepackage[caption=false,font=normalsize,labelfont=sf,textfont=sf]{subfig}
\usepackage{textcomp}
\usepackage{stfloats}
\usepackage{url}
\usepackage{graphicx}
\usepackage{balance}
\usepackage{orcidlink}
\hypersetup{hidelinks}
\usepackage{tikz}
\usepackage{amssymb}
\usepackage{cite}
\usetikzlibrary{arrows.meta, positioning, calc, fit, shapes.geometric, decorations.pathreplacing}
\begin{document}
\setlength{\textfloatsep}{6pt plus 1pt minus 2pt}

\setlength{\floatsep}{4pt plus 1pt minus 1pt}

\setlength{\intextsep}{5pt plus 1pt minus 1pt}

\setlength{\dbltextfloatsep}{6pt plus 1pt minus 2pt}

\setlength{\dblfloatsep}{4pt plus 1pt minus 1pt}

\setlength{\abovecaptionskip}{2pt}
\setlength{\belowcaptionskip}{1pt}
\renewcommand{\topfraction}{0.95}
\renewcommand{\dbltopfraction}{0.95}
\renewcommand{\textfraction}{0.05}
\renewcommand{\floatpagefraction}{0.90}
\renewcommand{\dblfloatpagefraction}{0.90}

\title{LLM-Assisted Hierarchical Interference Control for Satellite-Terrestrial Integrated Networks}

\author{
    \IEEEauthorblockN{Heying Cao\,\orcidlink{0009-0006-6008-6143}, Hao Yuan\,\orcidlink{0000-0002-5670-6171}, and Xing Zhang\,\orcidlink{0000-0003-4345-6166}, \IEEEmembership{Senior Member, IEEE}}%
\thanks{Heying Cao, Hao Yuan, and Xing Zhang are with the School of Information and Communication Engineering, Beijing University of Posts and Telecommunications, Beijing 100876, China (e-mail: heying@bupt.edu.cn; yuan\_hao@bupt.edu.cn; hszhang@bupt.edu.cn).}
}


\markboth{IEEE Wireless Communications Letters}%
{LLM-Assisted Hierarchical Interference Control}
\maketitle

\begin{abstract}
Time-varying aggregate interference from low-Earth-orbit (LEO) satellites challenges terrestrial-network (TN) protection, whereas conservative control degrades LEO quality of service (QoS). This letter
proposes a hierarchical framework combining geometry-based interference
estimation, QoS-constrained power control, adaptive energy detection,
large language model (LLM)-assisted candidate ranking, stay-or-switch
decisions, and target-band selection, with deterministic verification.
The current band is retained only when both spectrum-hole and
power-feasibility conditions hold, with periodic reevaluation. Over a 300-s horizon with 0.5-s sampling, the proposed method
keeps the aggregate INR from exceeding the $-12.2$~dB threshold
and achieves a time-averaged mean LEO-link rate of
$320.97$~Mbps with $100\%$ QoS satisfaction. It markedly
outperforms power-only and satellite-muting in jointly protecting
the TN and preserving LEO QoS.
\end{abstract}

\begin{IEEEkeywords}
LEO--TN coexistence, aggregate interference control, spectrum sensing, LLM.
\end{IEEEkeywords}

\section{Introduction}
\label{sec:intro}
\IEEEPARstart{S}{pectrum} sharing between low-Earth-orbit (LEO) satellite systems and terrestrial networks (TNs) improves spectrum utilization~\cite{Design,TimeDivision}, but also introduces cross-system interference. Due to orbital mobility and periodic visibility, the aggregate interference from LEO satellites to terrestrial receivers is highly time-varying and spatially non-uniform~\cite{ADynamic}. This creates a fundamental tradeoff: overly conservative protection wastes LEO communication resources, whereas insufficient control may violate the TN interference requirement.

Existing studies have investigated this problem from the perspectives of interference modeling, resource management, spectrum sensing, and intelligent decision-making. 
Stochastic geometry, in-band coexistence analysis, satellite selection, and time-elevation interference models provide useful analytical tools for evaluating interference and coexistence feasibility~\cite{CoExistence,Feasibility,Satelliteselection,ADynamic}. 
However, these studies do not fully exploit deterministic orbital geometry for timely, slot-level interference assessment and control. 
Beam hopping, frequency assignment, beam-footprint design, time-division sharing, and graph-theoretic multi-satellite coordination have also been used to mitigate cross-system resource conflicts and tilted-beam interference~\cite{TimeDivision,Beammanagement,Beamfootprint,ADistributedBeam,Multi-Satellite}. 
Such methods often require extensive coordination and combinatorial optimization, which may hinder real-time interference protection in rapidly varying scenarios. Energy detection and cooperative sensing support dynamic spectrum access by identifying occupied bands and potential spectrum holes~\cite{Efficient,Collaborative,SpectrumSensing}. 
Reinforcement learning and large language model (LLM)-aided methods provide new tools for complex resource-allocation decisions~\cite{Intelligent,LLMaided}. Nevertheless, existing intelligent methods often lack tight integration with orbital geometry and physical-layer interference constraints, limiting their direct applicability to safety-critical interference control.

Motivated by these observations, this letter proposes an LLM-assisted
hierarchical interference-control framework for LEO--TN coexistence.
It prioritizes current-band power control while monitoring spectrum
occupancy. Alternative spectrum holes are screened when current-band
operation is infeasible or unavailable, or when periodic reevaluation
is due. The LLM ranks the screened candidates and jointly determines
the stay-or-switch action and switching target, followed by deterministic
safety verification.

The main contributions are summarized as follows:
\begin{itemize}
    \item A hierarchical strategy coordinates QoS-constrained power
    control and adaptive spectrum sensing. Alternative bands are
    evaluated upon current-band infeasibility or unavailability and
    during periodic reevaluation, thereby limiting unnecessary switching.

    \item An adaptive-sensing-driven LLM mechanism ranks screened
    spectrum-hole candidates and jointly selects the stay-or-switch
    action and target band using sensing, INR/ACIR, and LEO QoS
    information. Deterministic verification enforces the system constraints.
\end{itemize}

\section{System Model and Problem Formulation}

\subsection{Terrestrial Network}
\label{sec:system}

Consider a TN modeled by a clustered binomial point process (BPP).
A total of $N_c$ TN base stations (BSs) are independently and uniformly distributed in a disk with radius $r_{\rm TN}$, and the protected TN receiver is located at an offset $x_0$ from the cluster center. 
The serving BS is the nearest BS. 
The TN link follows a power-law path-loss model with exponent $\alpha_{\rm TN}=3$ and Nakagami-$m$ fading ($m=1$, corresponding to Rayleigh fading), consistent with~\cite{CoExistence}.

\subsection{LEO Satellite System}

The LEO satellites follow a Walker constellation denoted by $N/P/F:h,i$~\cite{ADynamic}.
At time slot $t$, the visible satellite set is defined as
\begin{equation}
\mathcal{V}(t)=
\left\{k \mid \theta_k(t)\geq \theta_{\min}\right\},
\end{equation}
where $\theta_k(t)$ is the elevation angle of satellite $k$ observed from the protected TN receiver, and $\theta_{\min}$ is the minimum visible elevation angle.

For each visible satellite, the transmit beam is assumed to point toward the nadir direction. 
The off-axis angle $\phi_k(t)$ determines the satellite transmit antenna gain according to the ITU-R S.1528 pattern~\cite{Satelliteantenna}. 
The protected TN receiver uses an omnidirectional antenna ($0$ dBi). 
For a $2$~GHz carrier and $35.16$ dBi peak satellite gain, the corresponding $3$ dB beamwidth is approximately $\phi_b\approx3^\circ$.

\begin{figure}[!t]
\centering
\includegraphics[width=\columnwidth]{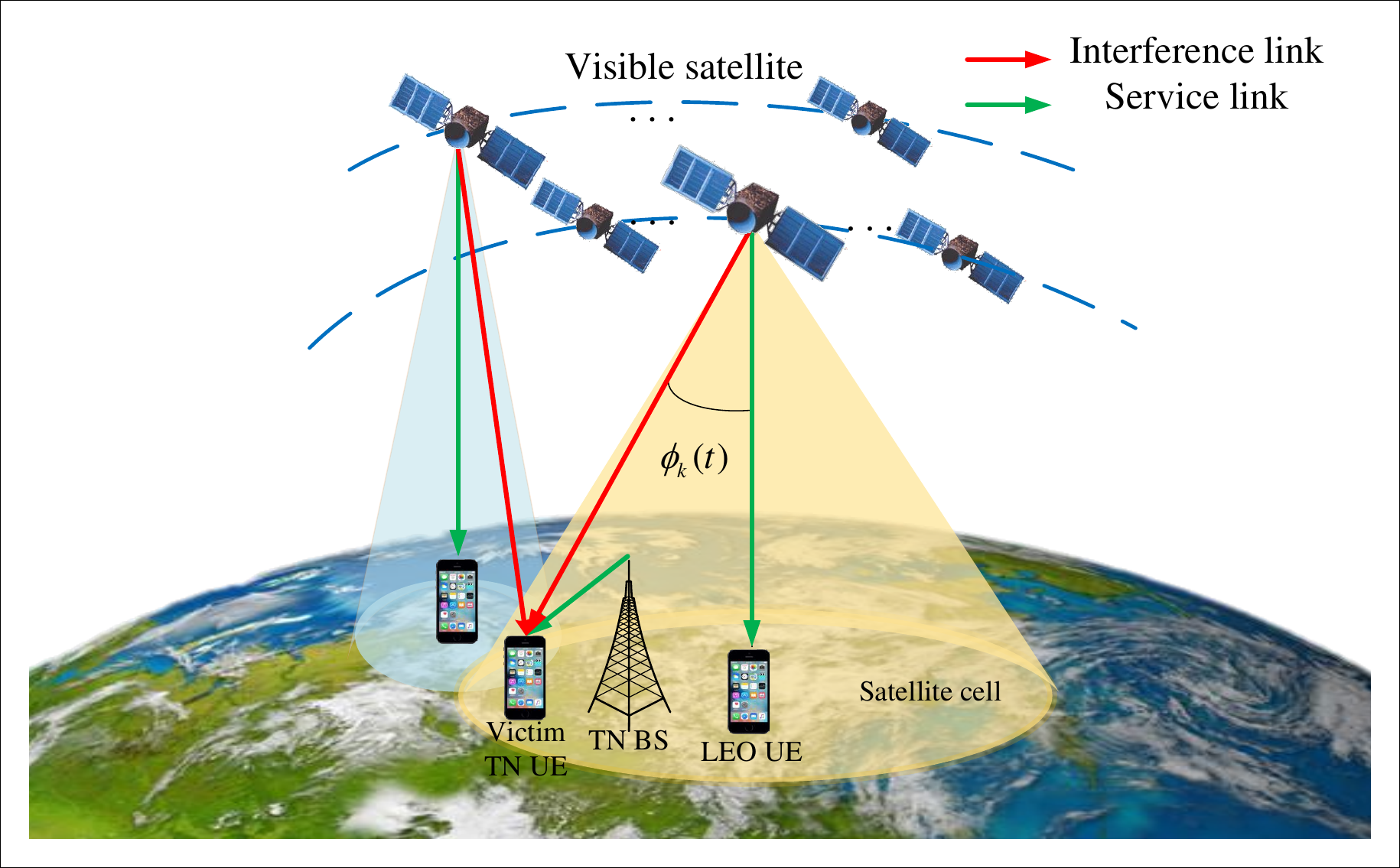}
\caption{Scenario of LEO--TN spectrum sharing and aggregate interference.}
\label{fig:scenario}
\end{figure}

\begin{figure}[!t]
\centering
\includegraphics[width=\columnwidth]{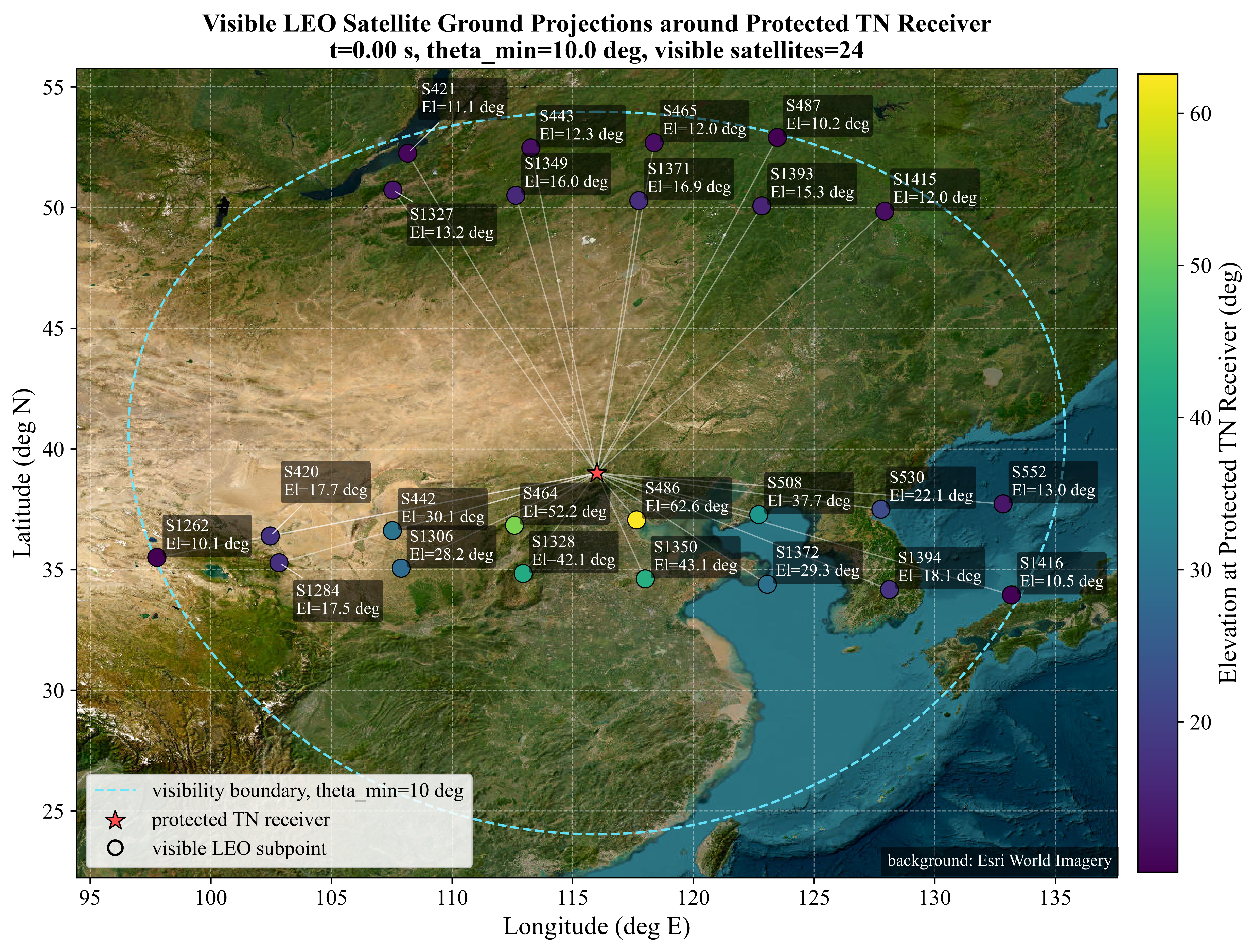}
\caption{Projected positions of visible LEO satellite subpoints for a protected TN receiver at \((39^\circ{\rm N},116^\circ{\rm E})\).}
\label{fig:visible_satellites}
\end{figure}

\subsection{Aggregate Interference}
\label{sec:aggregate_interference}

Let $p_k(t)$ be the transmit power of satellite $k$ at time slot $t$, with $p_k(t)\leq P_{\max}$. Let $d_k(t)$ denote the slant range from satellite $k$ to the protected TN receiver.
Considering free-space propagation, the channel gain from satellite $k$ to the protected TN receiver is
\begin{equation}
h_k^{\rm LEO\rightarrow TN}(t)
=
G_k^{\rm tx}\!\left(\phi_k(t)\right)G_{\rm TN}^{\rm rx}
\left(
\frac{\lambda}{4\pi d_k(t)}
\right)^2,
\end{equation}
where $\lambda=c/f_c$ is the wavelength and $G_{\rm TN}^{\rm rx}$ is the TN receive antenna gain.

Let $A_{k,j}^{\rm ACIR}(t)$ denote the applicable ACIR in dB. The associated
linear band-coupling factor is
$\rho_{k,j}(t)=10^{-A_{k,j}^{\rm ACIR}(t)/10}\in[0,1]$.
The band-dependent aggregate interference is
\begin{equation}
I_j^{\rm LEO\rightarrow TN}(t)
=
\sum_{k\in\mathcal{V}(t)}
\rho_{k,j}(t)p_k(t)h_k^{\rm LEO\rightarrow TN}(t).
\end{equation}
For co-channel operation, $\rho_{k,j}(t)=1$.

The maximum allowable aggregate interference is determined by the INR
threshold as
\begin{equation}
I_{\rm cap}
=
\sigma_{\rm TN}^2 10^{\Gamma_{\rm cap}/10},
\end{equation}
where $\sigma_{\rm TN}^2$ is the TN receiver noise power and
$\Gamma_{\rm cap}$ is the INR threshold in dB.

\subsection{Problem Formulation}

Let $\boldsymbol{p}(t)=[p_k(t)]_{k\in\mathcal{V}(t)}$ and $j(t)$
denote the satellite power vector and the selected band, respectively,
and let $j_{\rm c}(t)=j(t-1)$ denote the current operating band.
Let $\mathcal{F}$ denote the monitored band set, including the current
band, and let $\mathcal{F}_{\rm hole}(t)\subseteq\mathcal{F}$ denote
the bands declared as spectrum holes by energy detection.

The quantities $U_{\rm R}(t)$, $C_{\rm INR}(t)$, and $U_{\rm H}(t)$
denote the normalized aggregate-rate utility, aggregate-INR cost, and
spectrum-hole-quality utility of the selected configuration,
respectively.

Let $f_j$ denote the center frequency of band $j$. The normalized
band-change cost is defined as
\begin{equation}
C_{\rm sw}(t)
=
\frac{
\left|f_{j(t)}-f_{j_{\rm c}(t)}\right|
}{
\max_{a,b\in\mathcal{F}}
\left|f_a-f_b\right|
}.
\label{eq:switching_cost}
\end{equation}
It is zero when the current band is retained and increases with the
frequency distance of a reconfiguration.

The joint power--spectrum control problem is formulated as
\begin{subequations}
\label{prob:joint_control}
\begin{align}
\mathcal{P}(t):\quad
&\max_{\boldsymbol{p}(t),\,j(t)}
\quad
\begin{aligned}[t]
&\omega_{\rm R}U_{\rm R}(t)
-\omega_{\rm I}C_{\rm INR}(t)\\[-1mm]
&+\omega_{\rm H}U_{\rm H}(t)
-\omega_{\rm S}C_{\rm sw}(t)
\end{aligned}
\label{prob:objective}
\\
\mathrm{s.t.}\quad
&I_{j(t)}^{\rm LEO\rightarrow TN}(t)
\leq I_{\rm cap},
\label{eq:tn_constraint}
\\
&R_{k,j(t)}^{\rm LEO}(t)
\geq R_{\min},
\quad \forall k\in\mathcal{V}(t),
\label{eq:leo_qos_constraint}
\\
&0\leq p_k(t)\leq P_{\max},
\quad \forall k\in\mathcal{V}(t),
\label{eq:power_constraint}
\\
&j(t)\in\mathcal{F}_{\rm hole}(t).
\label{eq:spectrum_action_constraint}
\end{align}
\end{subequations}

Here, $\omega_{\rm R}$, $\omega_{\rm I}$, $\omega_{\rm H}$, and
$\omega_{\rm S}$ are nonnegative weights. Constraints
\eqref{eq:tn_constraint}--\eqref{eq:spectrum_action_constraint}
enforce TN protection, LEO QoS, transmit-power limits, and
spectrum-hole selection, respectively. Constraint
\eqref{eq:spectrum_action_constraint} excludes bands detected as
occupied, whereas $U_{\rm H}(t)$ distinguishes the sensing quality
among the remaining spectrum holes. The proposed method provides a
low-complexity heuristic solution to~\eqref{prob:joint_control}
without claiming global optimality.

\section{Proposed Method}
\label{sec:method}

Algorithm~\ref{alg:hierarchical_control} summarizes the hierarchical
heuristic guided by Problem~\eqref{prob:joint_control}. The controller
first evaluates the current band and screens alternative spectrum holes
when required. The LLM ranks the screened candidates and jointly selects
the stay-or-switch action and target band. Deterministic power control
and safety verification precede execution.

\begin{algorithm}[!t]
\caption{Hierarchical Power--Spectrum Control}
\label{alg:hierarchical_control}
\footnotesize
\renewcommand{\algorithmicrequire}{\textbf{Input:}}
\renewcommand{\algorithmicensure}{\textbf{Output:}}
\begin{algorithmic}[1]
\REQUIRE Current band $j_{\rm c}$, band set $\mathcal{F}$, and
epoch-$t$ observations
\ENSURE A feasible $(\boldsymbol{p}(t),j(t))$ or \textsc{Blocked}
\STATE \textbf{Current band:}
$(\boldsymbol{p}^{({\rm c})},q_{\rm c})\leftarrow
\textsc{BandPower}(j_{\rm c})$
\STATE $(a_{\rm c},r_{\rm per})\leftarrow
(\textsc{IsHole}(j_{\rm c}),\textsc{PeriodicDue}(t))$
\IF{$q_{\rm c}=1\land a_{\rm c}=1\land r_{\rm per}=0$}
    \RETURN $(\boldsymbol{p}^{({\rm c})},j_{\rm c})$
\ENDIF
\STATE \textbf{Alternatives:}
$\mathcal{C}\leftarrow\textsc{ScreenRank}
(\textsc{Sense}(\mathcal{F})\setminus\{j_{\rm c}\})$
\IF{$\mathcal{C}=\emptyset$}
    \RETURN $(\boldsymbol{p}^{({\rm c})},j_{\rm c})$ if
    $q_{\rm c}=1$; otherwise \textsc{Blocked}
\ENDIF
\STATE \textbf{LLM decision:}
$(a,j^\star)\leftarrow\textsc{LLMDecide}
(j_{\rm c},q_{\rm c},\mathcal{C})$
\IF{$a=\mathrm{stay}\land q_{\rm c}=1$}
    \RETURN $(\boldsymbol{p}^{({\rm c})},j_{\rm c})$
\ELSIF{$a=\mathrm{stay}$}
    \STATE $j^\star\leftarrow\textsc{First}(\mathcal{C})$
    \COMMENT{Override unsafe stay}
\ENDIF
\IF{$a=\mathrm{switch}\land q_{\rm c}=1\land a_{\rm c}=1
\land r_{\rm per}=1
\land\Delta S(t)<\delta_{\rm re}$}
    \RETURN $(\boldsymbol{p}^{({\rm c})},j_{\rm c})$
    \COMMENT{Periodic hysteresis}
\ENDIF
\STATE $\mathcal{L}\leftarrow[j^\star]\oplus
(\mathcal{C}\setminus\{j^\star\})$
\FOR{each $j\in\mathcal{L}$}
    \STATE $(\boldsymbol{p}^{(j)},q_j)\leftarrow
    \textsc{BandPower}(j)$
    \IF{$q_j=1\land\textsc{Verify}(\boldsymbol{p}^{(j)},j)$}
        \RETURN $(\boldsymbol{p}^{(j)},j)$
    \ELSIF{$q_{\rm c}=1$}
        \RETURN $(\boldsymbol{p}^{({\rm c})},j_{\rm c})$
    \ENDIF
\ENDFOR
\RETURN \textsc{Blocked}
\end{algorithmic}
\end{algorithm}
\subsection{Current-Band Power Control}

Let $j_{\rm c}(t)=j(t-1)$ denote the current operating band. At the beginning
of slot $t$, the aggregate interference under maximum
satellite transmit power is
\begin{equation}
I_0(t)
=
\sum_{k\in\mathcal{V}(t)}
\rho_{k,j_{\rm c}(t)}(t)P_{\max}
h_k^{\rm LEO\rightarrow TN}(t).
\label{eq:initial_interference}
\end{equation}
If $I_0(t)\leq I_{\rm cap}$ and the maximum-power configuration satisfies
the LEO QoS constraints, the current band is retained.
Otherwise, QoS-aware power control is performed on the current band.

From the Shannon rate constraint, the minimum transmit power required to satisfy the LEO QoS requirement $R_{\min}$ is given by
\begin{equation}
p_{k,j}^{\rm QoS}(t)
=
\frac{
\left(2^{R_{\min}/B_{\rm LEO}}-1\right)
\left[
\sigma_{\rm LEO}^2+
I_{k,j}^{\rm TN\rightarrow LEO}(t)
\right]
}{
h_k^{\rm LEO}(t)
}.
\label{eq:qos_power}
\end{equation}
Thus, power control is constrained by
\begin{equation}
p_{k,j}^{\rm QoS}(t)\leq p_k(t)\leq P_{\max}.
\end{equation}

The visible satellites are sorted according to their interference contributions, and the transmit powers of dominant interferers are reduced first. 
If the TN protection constraint can be satisfied without violating the QoS-aware lower bounds, no frequency switching is required. 
Otherwise, spectrum sensing is activated.

\subsection{Adaptive Spectrum Sensing}

When current-band power control is infeasible, the alternative bands in the sensing set $\mathcal{F}$ are examined.
For band $j$ at time slot $t$, the received sensing signal is
\begin{equation}
y_j[n,t]
=
s_j^{\rm TN}[n,t]+w_j[n,t],
\end{equation}
where $s_j^{\rm TN}[n,t]$ is the TN signal on band $j$, and $w_j[n,t]\sim\mathcal{CN}(0,\sigma_w^2)$ is AWGN.

The binary hypotheses are
\begin{equation}
\begin{cases}
\mathcal{H}_0: y_j[n,t]=w_j[n,t],\\
\mathcal{H}_1: y_j[n,t]=s_j^{\rm TN}[n,t]+w_j[n,t].
\end{cases}
\end{equation}

Let \(P_{{\rm TN},j}\triangleq
\mathbb{E}[|s_j^{\rm TN}[n,t]|^2]\) denote the received TN signal
power at the sensing receiver on band \(j\). The sensing SNR is defined as
\begin{equation}
\gamma_{{\rm sen},j}
=
10\log_{10}
\left(
\frac{P_{{\rm TN},j}}{\sigma_{w,j}^{2}}
\right).
\label{eq:sensing_snr}
\end{equation}
Accordingly, an occupied band with prescribed sensing SNR
\(\gamma_{\rm sen}\) has
\(P_{{\rm TN},j}=\sigma_{w,j}^{2}10^{\gamma_{\rm sen}/10}\).
This quantity characterizes TN-activity detectability at the sensing
receiver and is distinct from the LEO-to-TN INR constraint.

The energy statistic is
\begin{equation}
E_j(t)
=
\frac{1}{N}
\sum_{n=1}^{N}
|y_j[n,t]|^2.
\end{equation}

The background energy of each band is estimated from a trusted $\mathcal{H}_0$ sliding window. 
Let $\hat{\mu}_{0,j}(t)$ and $\hat{\sigma}_{0,j}(t)$ denote the estimated background mean and standard deviation. 
Given a target false-alarm probability $P_{\rm fa}^{\rm tar}$, the adaptive threshold is
\begin{equation}
\lambda_j(t)
=
\hat{\mu}_{0,j}(t)
+
Q^{-1}
\left(
P_{\rm fa}^{\rm tar}
\right)
\hat{\sigma}_{0,j}(t).
\end{equation}

The sensing rule is
\begin{equation}
E_j(t)
\mathop{\gtrless}_{\mathcal{H}_0}^{\mathcal{H}_1}
\lambda_j(t).
\end{equation}
Only samples detected as $\mathcal{H}_0$ are used to update the background window, while samples detected as $\mathcal{H}_1$ are excluded to avoid threshold chasing. 
The available spectrum-hole set is
\begin{equation}
\mathcal{F}_{\rm hole}(t)
=
\left\{
j\in\mathcal{F}
\mid
E_j(t)\leq\lambda_j(t)
\right\}.
\end{equation}

\subsection{Prompt-Engineered LLM-Assisted Spectrum Decision}

\begin{figure*}[!t]
\centering
\includegraphics[width=0.82\textwidth]{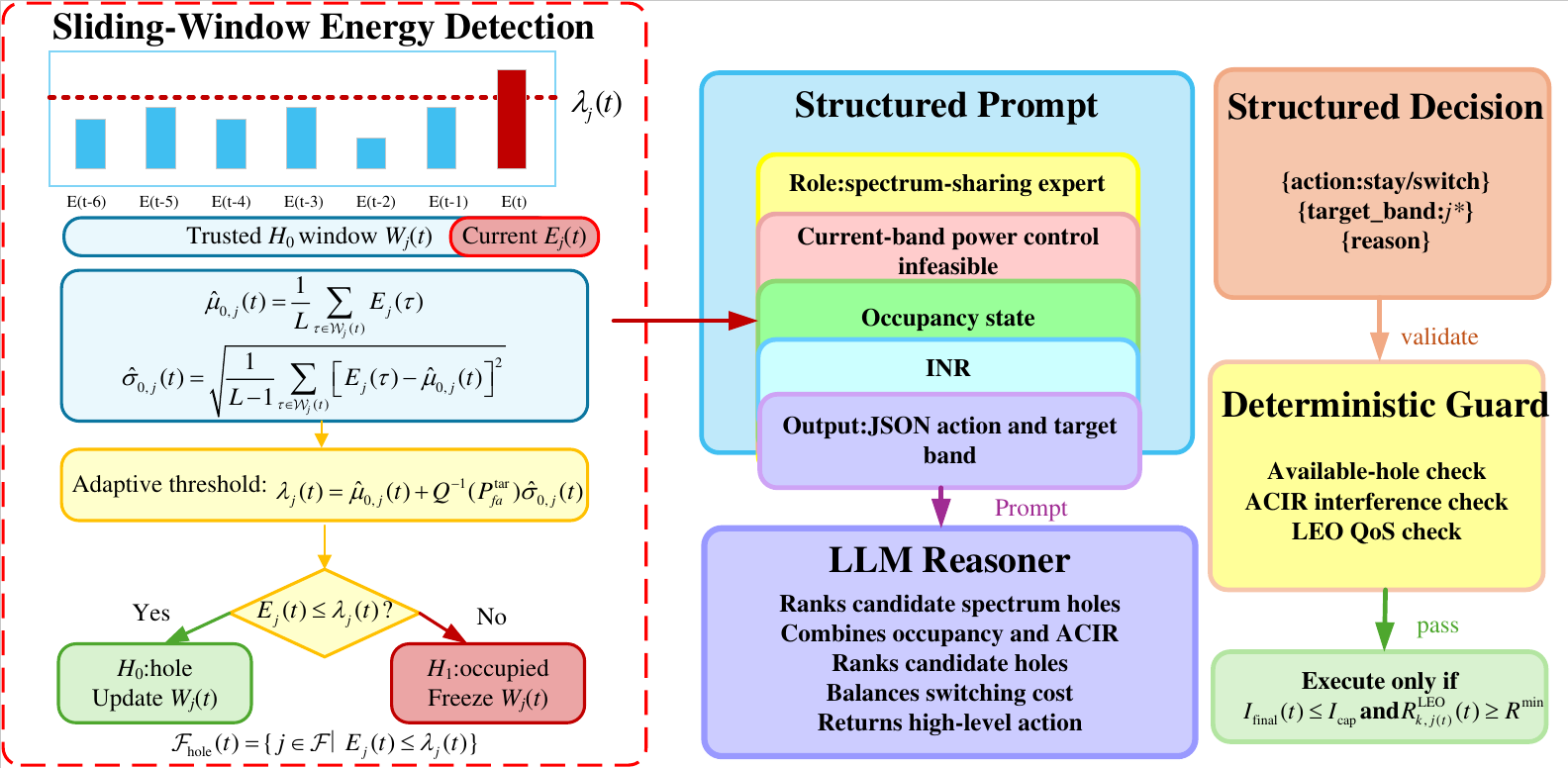}
\caption{Flowchart of the proposed LLM-assisted hierarchical interference-control procedure.}
\label{fig:agent_flow}
\end{figure*}

As illustrated in Fig.~\ref{fig:agent_flow}, the LLM-assisted module performs prompt-engineered spectrum selection.
The prompt follows a structured design inspired by LLM-aided spectrum allocation and consists of four parts.

\textit{Domain constraint:} 
The LLM is prompted as a satellite-terrestrial spectrum-sharing expert.
It is instructed to select a safe spectrum action that protects the TN receiver, maintains LEO QoS, and avoids unnecessary frequency switching.

\textit{Variable declaration:} 
The prompt declares the current band, candidate bands, aggregate-INR estimates, TN protection threshold, power-control feasibility, LEO QoS status, spectrum sensing results, ACIR-based interference estimates, and switching cost.

\textit{Task description:} 
The LLM is asked to choose between staying on the current band and switching to a candidate spectrum hole.
Occupied bands are prohibited. 
If power control is infeasible and the current band is unsafe, the stay action is not allowed.

\textit{Real-time observation and output:} 
Slot-level sensing and interference-estimation results are inserted into the prompt, and the LLM outputs a structured decision with the action, target band, risk label, and a short explanation.
The output is then passed to the deterministic safety-verification module before execution.

\subsection{Safety Verification}

Before execution, the selected action is checked by deterministic safety verification. 
Each ranked candidate is accepted only if all constraints in
Problem~\eqref{prob:joint_control} are jointly satisfied.
If the selected band fails verification, other candidate bands are tested; if no candidate satisfies these constraints, the current slot is marked as blocked.

\section{Simulation Results}
\label{sec:exp}

\subsection{Experimental Setup}

Table~\ref{tab:simulation_parameters} summarizes the main parameters. The LLM-assisted module uses DeepSeek-R1:7B with deterministic safety verification.

\textbf{Baselines:} \emph{No control} uses $P_{\max}$ on the reference
band. \emph{Power-only} keeps the band fixed and reduces power, disabling
satellites only when necessary. \emph{Satellite muting} keeps the band
fixed and successively disables the strongest interferers while the
others transmit at $P_{\max}$.

\begin{table}[!t]
\centering
\caption{Main Simulation Parameters}
\label{tab:simulation_parameters}
\setlength{\tabcolsep}{4pt}
\begin{tabular}{@{}l c c@{}}
\hline
Parameter & Symbol & Value \\
\hline
LEO constellation
& / & Walker $72\times 22$ \\
Orbital altitude
& $h$ & $550$~km \\
Orbital inclination
& $i$ & $80^\circ$ \\
Minimum visible elevation
& $\theta_{\min}$ & $10^\circ$ \\
System bandwidth
& $B_{\rm LEO}$ & $20$~MHz \\
Observation window
& / & $5$~min \\
Slot duration
& / & $0.5$~s \\
Maximum transmit power
& $P_{\max}$ & $40$~W \\
LEO QoS requirement
& $R_{\min}$ & $100$~Mbps \\
TN INR threshold
& $\Gamma_{\rm cap}$ & $-12.2$~dB \\
Candidate frequency bands
& $|\mathcal{F}|$ & $9$ \\
Sensing setup
& $(\gamma_{\rm sen},N,L_{\rm bg})$
& $(30$~dB$,1000,32)$ \\
False-alarm target
& $P_{\rm fa}^{\rm tar}$
& $0.01$ \\

\hline
\end{tabular}
\end{table}
\vspace{-0.35\baselineskip}
\subsection{Aggregate INR Performance}
\begin{figure}[!t]
\centering
\includegraphics[width=0.72\columnwidth]{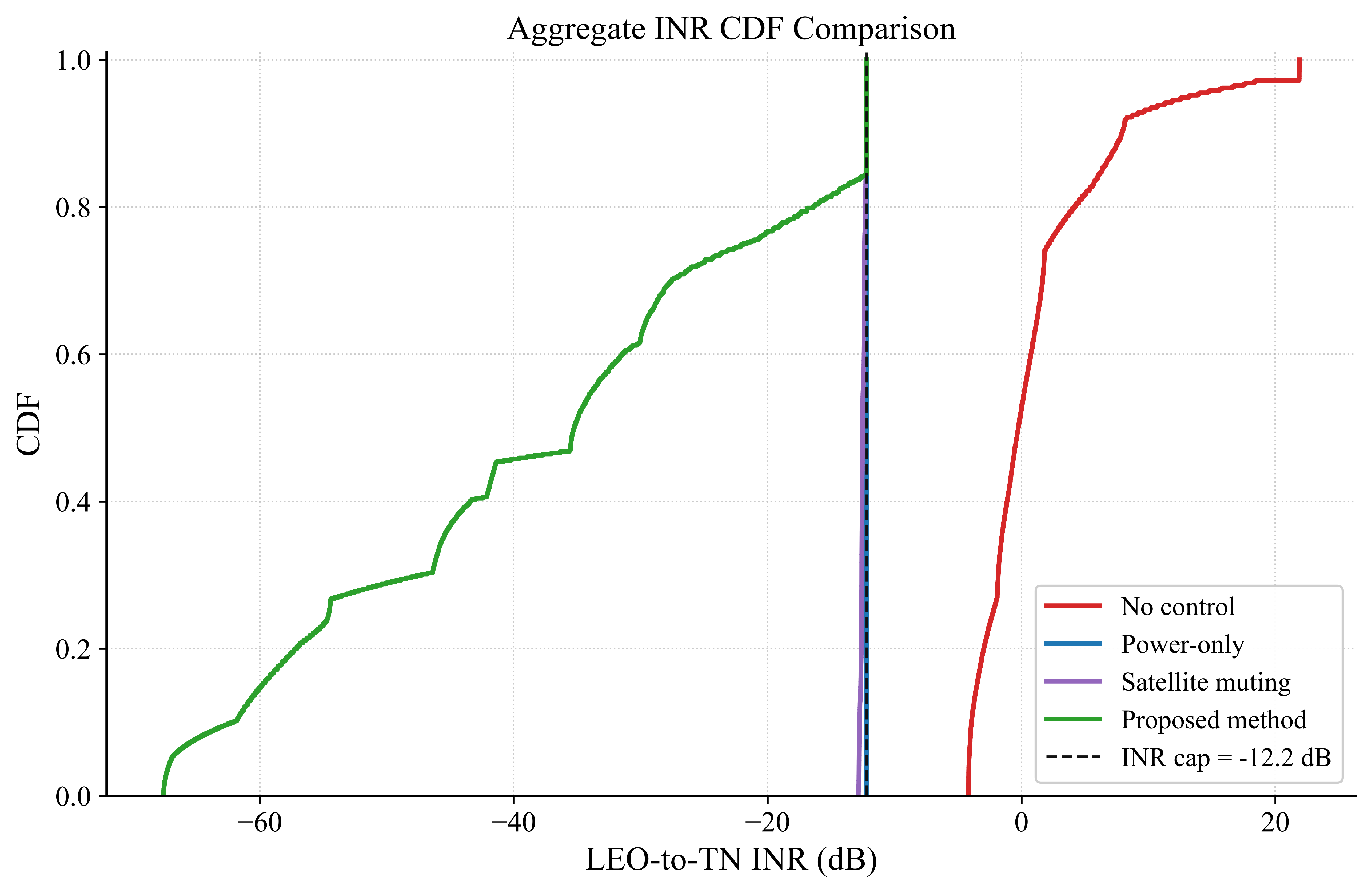}
\caption{Aggregate-INR CDFs under the four schemes.}
\label{fig:main_inr_cdf}
\end{figure}

Fig.~\ref{fig:main_inr_cdf} shows the aggregate-INR CDF. No control exceeds the threshold in almost all slots. Power-only and satellite muting suppress INR by reducing or disabling transmission. The proposed method keeps the INR below the $-12.2$~dB threshold.

\begin{figure}[!t]
\centering
\includegraphics[width=\columnwidth]{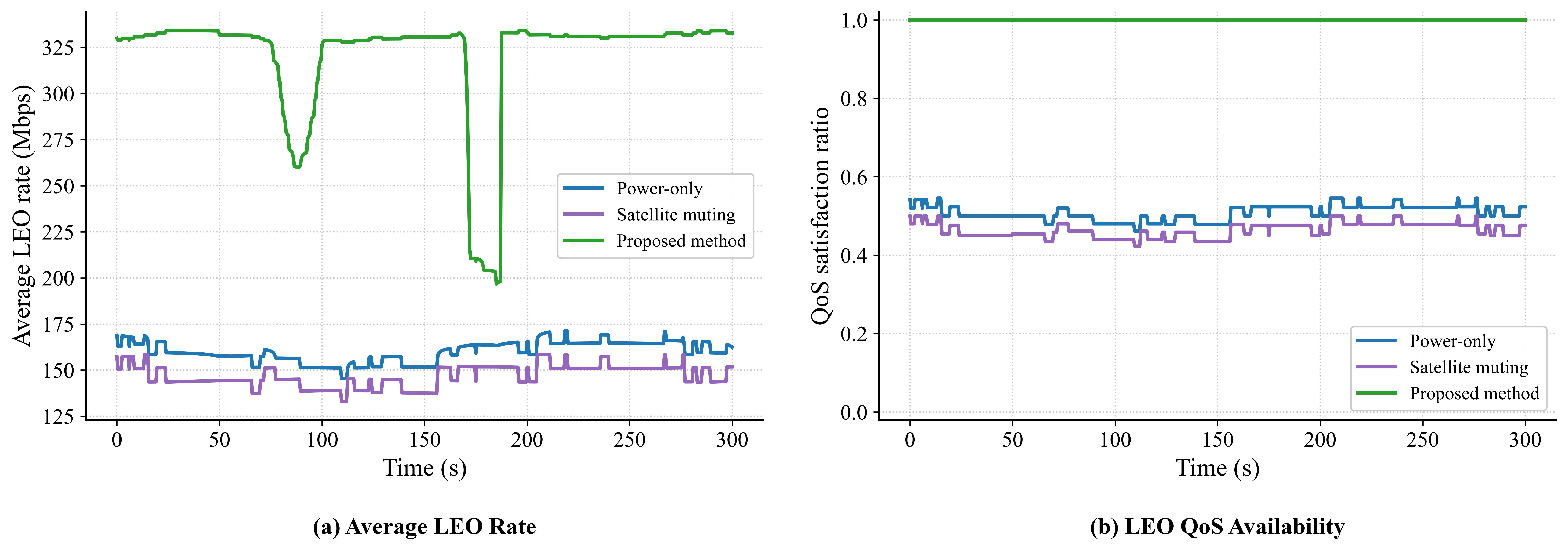}
\caption{LEO QoS comparison under different interference-mitigation schemes.}
\label{fig:qos_comparison}
\end{figure}

Fig.~\ref{fig:qos_comparison} compares LEO rates. The proposed method achieves $320.97$~Mbps with $100\%$ QoS satisfaction. Power-only and muting reduce the rate to $159.97$ and $146.89$ Mbps, with QoS satisfaction of $50.8\%$ and $46.4\%$, respectively.
Therefore, directly suppressing interference by power reduction or muting can protect the TN receiver, but it significantly sacrifices LEO service quality.

\vspace{-0.35\baselineskip}
\subsection{Spectrum Sensing and Frequency Selection}
\label{sec:comparison}

\begin{figure}[!t]
\centering
\includegraphics[width=\columnwidth]{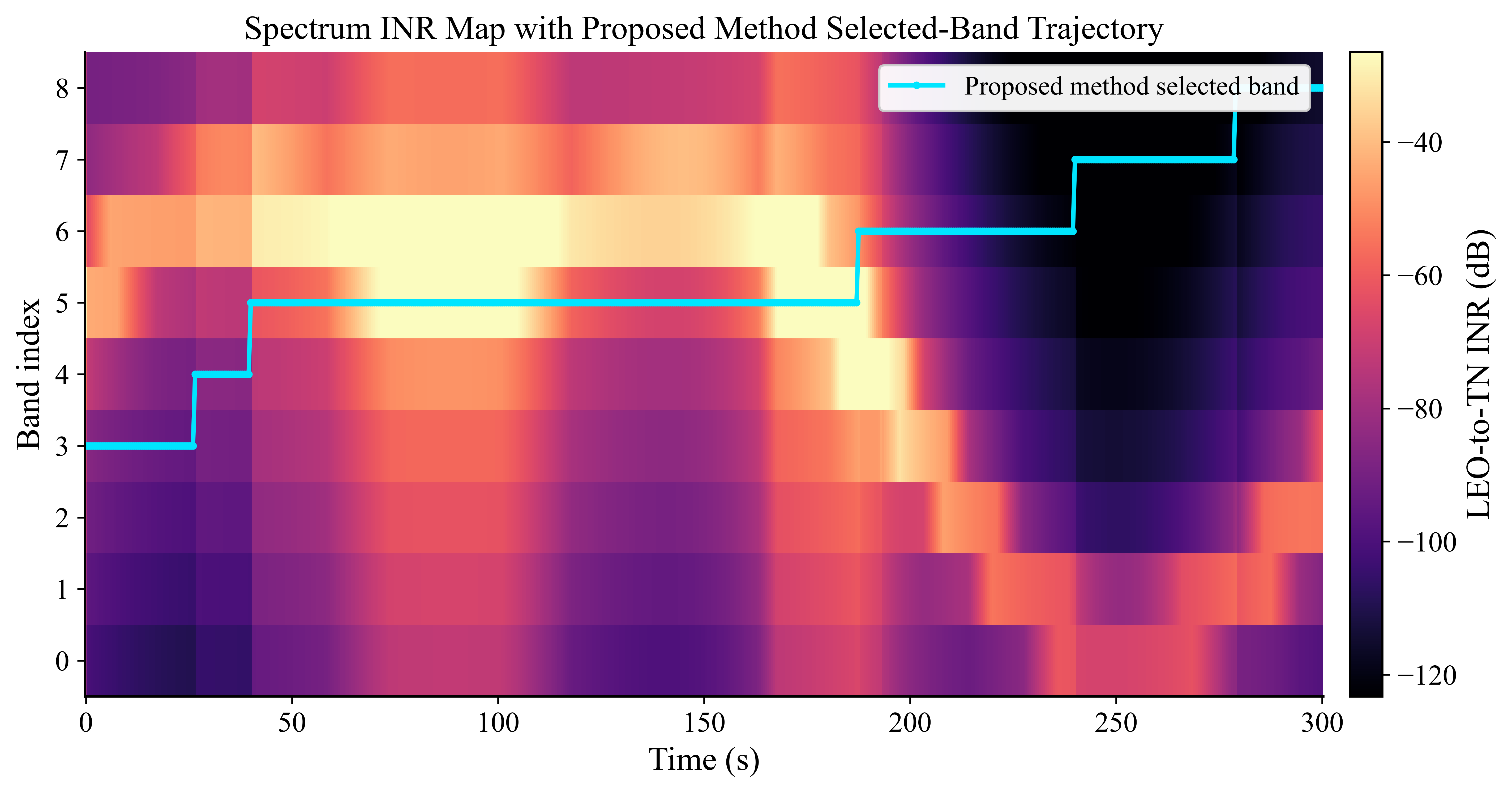}
\caption{Time-frequency INR map and selected-band trajectory of the proposed method.}
\label{fig:spectrum_heatmap}
\end{figure}

Fig.~\ref{fig:spectrum_heatmap} presents the time-frequency INR map with the selected-band trajectory. 
The INR distribution varies over time because the dominant interfering satellite and the corresponding ACIR reference band change with orbital movement. 
The selected trajectory shows that the proposed method selects a band that satisfies the TN protection constraint while considering sensing availability and switching cost, rather than always choosing the minimum-INR band.

\vspace{-0.35\baselineskip}
\subsection{Sensitivity Analysis}
\begin{figure}[!t]
\centering
\includegraphics[width=\columnwidth]{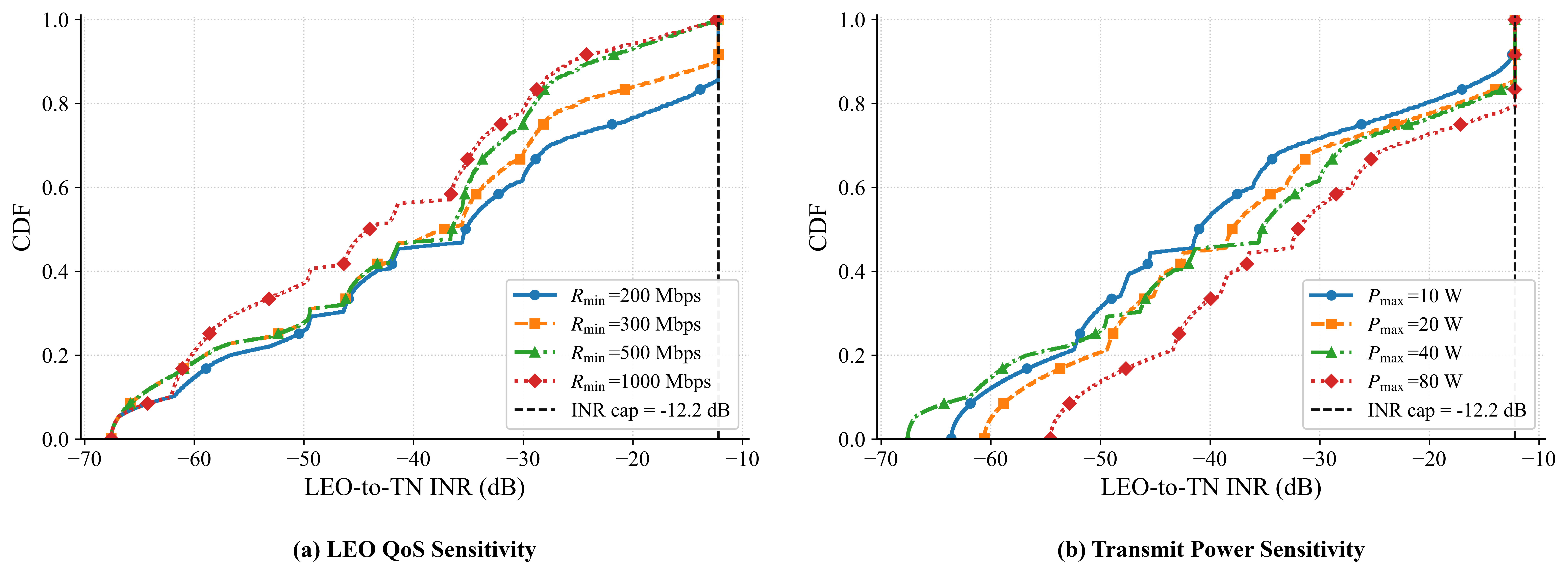}
\caption{Sensitivity analysis of the proposed method: (a) LEO QoS requirement; (b) maximum satellite transmit power.}
\label{fig:pmax_and_qos_sensitivity}
\end{figure}

Fig.~\ref{fig:pmax_and_qos_sensitivity} evaluates the effects of
$R_{\min}$ and $P_{\max}$. A larger $R_{\min}$ raises the QoS-aware power lower bound and may increase the need for spectrum switching.As $P_{\max}$ increases, the potential aggregate interference becomes stronger, making TN protection more challenging.
In both cases, the proposed method keeps the INR below the protection threshold, demonstrating its robustness to different QoS and transmit-power settings.

\vspace{-0.35\baselineskip}
\subsection{Energy-Detection Characterization}

\begin{figure}[!t]
\centering
\includegraphics[width=0.72\columnwidth]{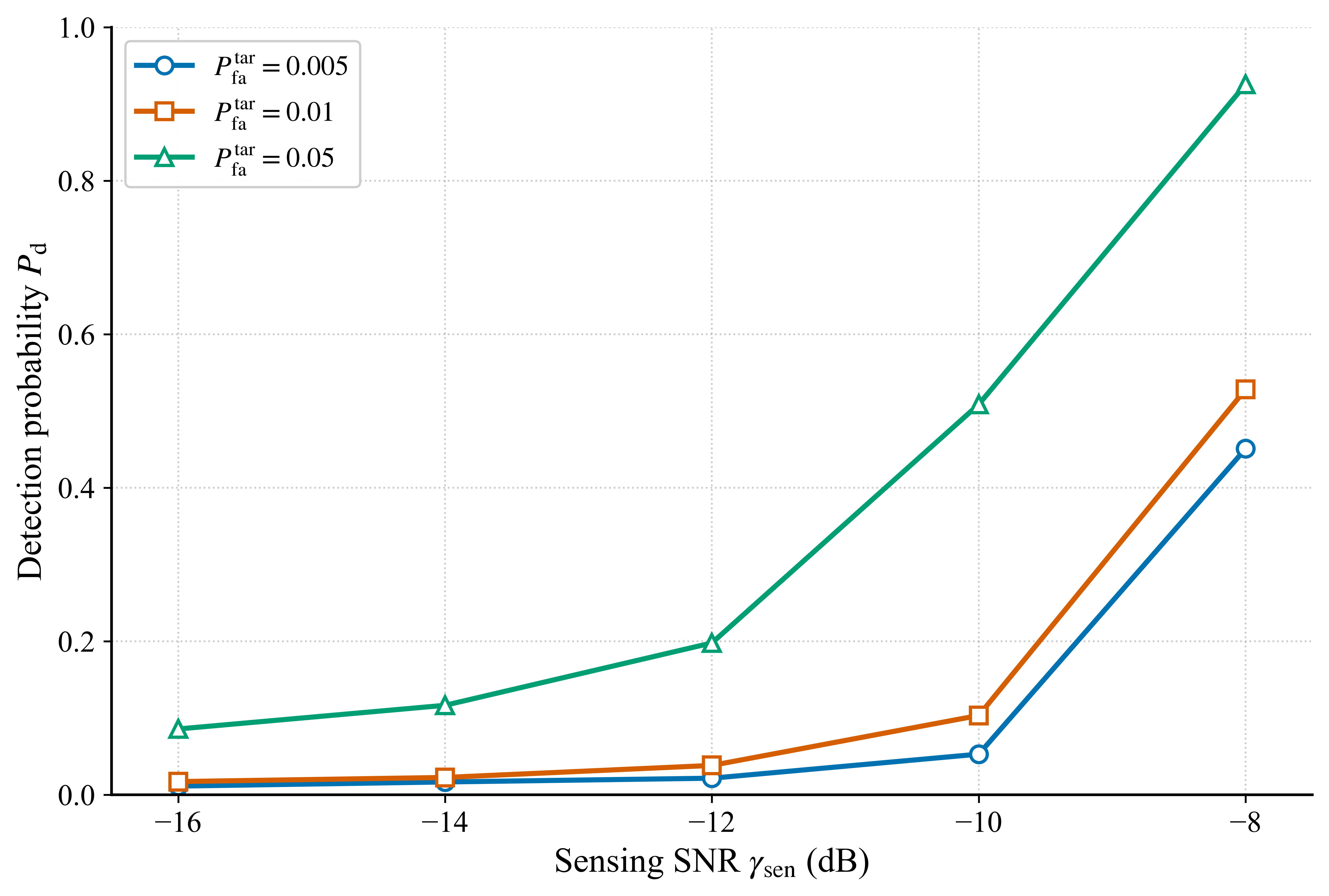}
\caption{Detection probability versus sensing SNR for three target false-alarm probabilities.}
\label{fig:ed_snr}
\end{figure}

Fig.~\ref{fig:ed_snr} characterizes detector performance over
sensing SNR for three false-alarm targets. 
The detection probability
increases with both $\gamma_{\rm sen}$ and $P_{\rm fa}^{\rm tar}$. For example, when
\(P_{\rm fa}^{\rm tar}=0.01\), \(P_{\rm d}\) rises from \(1.71\%\) at
\(\gamma_{\rm sen}=-16\)~dB to \(52.81\%\) at \(-8\)~dB. This demonstrates the low-SNR sensing tradeoff and the need for
deterministic safety verification.

\section{Conclusion}
\label{sec:conclusion}

This letter proposed an LLM-assisted hierarchical interference-control
framework for LEO--TN coexistence. By combining geometry-based
interference assessment, QoS-aware power control, adaptive spectrum
sensing, constrained spectrum selection, and deterministic safety
verification, the method balances TN protection, LEO QoS, and
switching overhead.


\bibliographystyle{IEEEtran}
\bibliography{references}
\end{document}